\documentstyle[12pt]{article}

\def\be{\begin{equation}}
\def\ee{\end{equation}}
\def\bea{\begin{eqnarray}}
\def\eea{\end{eqnarray}}
\def\ba{\begin{array}}
\def\ea{\end{array}}
\def\pa{\partial}

\def\nn{\nonumber}

\def\si{\sigma}

\def\de{\delta}
\def\ka{\kappa}

\def\la{\lambda}
\def\La{\Lambda}
\def\eps{\epsilon}

\begin{document}
\hfill{IFT-20/2001}

\begin{center}
{\large\bf Brane localization of gravity\\
in higher derivative theory}\\
\vspace{8mm}
{\bf Krzysztof A. Meissner${}^a$ and Marek Olechowski${}^{a,b}$}\\
\vspace{4mm}
${}^a$Institute of Theoretical Physics, Warsaw University\\
Ho\.za 69, 00-681 Warsaw, Poland\\
\vspace{4mm}
${}^b$Physikalisches Institut der Universit\"at Bonn\\
Nussallee 12, 53115 Bonn, Germany
\end{center}
\begin{abstract}
We consider a class of higher order corrections in the form of Euler
densities of arbitrary rank $n$ to the standard gravity action in $D$
dimensions. We have previously shown that this class of
corrections allows for domain wall solutions despite the presence of
higher powers of the curvature. In the present paper we explicitly
solve the linearized equation of motion for gravity fluctuations
around the domain wall background and
show that there always exist one massless state (graviton)
propagating on the wall and a continuous tower of massive states 
propagating in the bulk. 
\end{abstract}

\newpage

It may very well be that the theory of gravity as we know it today is
only an effective theory and the usual Einstein--Hilbert action should be
supplemented with corrections involving higher powers of the curvature
tensor. This point of view is supported for example by string theory or
the presence of the conformal anomalies in all quantum field theories
coupled to gravity.

In this paper we are interested in the corrections of special type --
Euler densities of arbitrary order $n$ ($n$ being a power of the
curvature tensor) \cite{L} in arbitrary space--time dimension $D$. It was
shown in \cite{KMO} that in the presence of arbitrary number of Euler
densities in the lagrangian, there always exist domain wall solutions.
The order $n$ should be less or equal to $D/2$ where $D$ is the dimension
of space--time ($D=2n$
although formally total derivative gives part of the conformal anomaly).
Euler densities appear for example in the $\alpha'$ expansion of the
string theory effective action \cite{strings}.
The Euler density of the order $n=2$ (equal in this case to the
Gauss--Bonnet combination) has been discussed in the presence of
branes in many papers, some of which are \cite{KKL}--\cite{NN}.

In the present paper we analyze the linearized equations of motion for
the fluctuations of the metric around the domain wall solutions in the
theory with arbitrary number of Euler densities. It is a
generalizations of the original idea \cite{RS2} and the analysis performed
for Gauss--Bonnet $n=2$ case \cite{KKL}
(the Gauss--Bonnet term for intersecting domain walls has been
recently discussed in \cite{NN}). The fluctuations are assumed to be
graviton--like i.e. transverse and traceless.

It turns out that in spite of the presence of higher
powers of curvature the picture is very similar to the usual
Randall--Sundrum scenario \cite{RS2}. There exists one normalizable
massless bound state and a continuous tower of massive states with
small amplitude on the wall.
The strategy adopted to calculate the equations of motions is based on the
explicit formulae
for the Euler densities derived in our previous paper \cite{KMO}.

The metric of the domain wall is conformally flat (and this fact was
extensively used in \cite{KMO}) but this is no longer the
case when we add a fluctuation to the metric. Therefore calculating the
equations of motion for the fluctuation requires calculating a first order
correction (linear in the Weyl tensor) to the formulae in \cite{KMO}. When
adding all the contributions, the resulting equation of motion turns out
to be almost identical to the lowest order case, the only difference being
in the actual value of the coefficients of the equations. It  is extremely
important to notice that the graviton (normalizable massless mode confined
to the wall) exists independently of the number of dimensions and 
presence of the Euler
densities of higher order. It is quite amazing that the picture in
presence of Euler densities is almost identical to the lowest order
scenario \cite{RS2}.

The Euler densities in $D$ dimensions are defined (in the form notation) as
\be
\mbox{\boldmath$I$}^{(n)}=\frac1{(D-2n)!}
{\eps_{a_1a_2\cdots a_D}}
R^{a_1a_2}\wedge \cdots \wedge R^{a_{2n-1}a_{2n}}\nn\\
\wedge
e^{a_{2n+1}}\wedge \cdots \wedge e^{a_D}
\,.
\label{eq:I}
\ee
(For $D=2n$ they are topological invariants and formally total
derivatives but a careful regularization shows that one cannot discard them
neither in the action nor in the equations of motion since they
correspond to the conformal anomaly).

We will consider models in which $D$-dimensional gravitational
interactions are described by the sum of such Euler densities and in
which there is a $(D-2)$--brane (a domain wall).
The action is the sum of the bulk and brane contributions:
\bea
S
&=&
S_{\rm bulk}+S_{\rm brane}
\,,
\nn\\
S_{\rm bulk}
&=&
\int d^Dx \sqrt{-g} \sum_{n=0}^{n_{max}} \ka_n I^{(n)}
\,,
\label{eq:S}
\\
S_{\rm brane}
&=&
\int d^{D-1}x \sqrt{-\tilde g} \left(-\la + \la_1 {\tilde R}+\ldots \right)
\nn
\,.
\eea
The metric on the brane is given by
${\tilde g}_{\mu\nu}(x^\rho)=g_{\mu\nu}(x^\rho,y=0)$
where $y=x^D$ and
$\mu,\nu,\ldots=1,\ldots,D-1$ while $M,N,\ldots=1,\ldots,D$.
In the brane action we write explicitly only the most important term
-- the brane cosmological constant.
The first two terms of the bulk action are known from conventional
gravity. The one with $n=0$ corresponds to
the cosmological constant: $I^{(0)}=1$, $\ka_0=-\La$.
The one with $n=1$ is the usual Hilbert--Einstein term,
$I^{(1)}=R$ with the coefficient $\ka_1=(2\ka^2)^{-1}$.
The maximal number of the higher order terms is
$n_{max} \le [D/2]$ as discussed in \cite{KMO}.

We assume in the following that $\la_1=0$; $\la_1\ne 0$ would give 
additional contribution $\la_1 m^2 h_m(0)/2$ to the right hand side of 
(\ref{dely}) discussed later so it would affect only the massive modes.

Let us start with the bulk equations of motion. They  can be obtained
from the variation of the vielbein in the bulk action
(for $n \le (D-1)/2$)
\be
\sum_n\frac{\ka_n(D-2n)}{(D-2n)!}
\eps_{a_1a_2\cdots a_{D-1}a} R^{a_1a_2}\wedge
\cdots \wedge R^{a_{2n-1}a_{2n}}
\wedge e^{a_{2n+1}}\wedge \cdots
\wedge e^{a_{D-1}}=0.
\label{eqofm}
\ee
We can write the curvature two--form as
\be
R^{ab}=C^{ab}+\frac1{D-2}(e^a\wedge K^b-e^b\wedge K^a)
\label{eq:R}
\ee
where $C^{ab}$ is a two--form composed of the Weyl tensor
$C_{MNRS}$ while $K^a$ is a one--form which will play important role
in our calculations and which is defined (for invertible
vielbeins $e_M^a$) as
$K^a=K_{MN}e^{M a}dx^{N}=K^a{}_b e^b$ with
\be
K_{MN} = R_{MN}- \frac1{2(D-1)}g_{MN}R
\,.
\ee

The main purpose of this paper is to analyse localization of the
effective brane gravity in higher order theories proposed in our
previous paper \cite{KMO}. In order to do this
we will look for solutions of the linearized equations of motion
(\ref{eqofm}) for the fluctuations of the metric:
\be
g_{MN}=g_{MN}^0+\eps h_{MN}
\label{gfluct}
\ee
for which the background metric $g_{MN}^0$ is of the
domain wall type that was proven in \cite{KMO} to satisfy the
equations of motion (\ref{eqofm}). The line element of this domain
wall background is equal
\be
ds^2=e^{-2f(y)} \eta_{\mu\nu} dx^\mu dx^\nu + dy^2
\label{ds2}
\ee
with the warp factor function given by $f(y)=\si |y|$.
The fluctuation $h_{MN}$ is assumed to
propagate along the brane and be transverse and traceless:
\be
h_{DM}=0,
\ \ \ \ \
\eta^{\mu\nu}h_{\mu\nu}=k^\mu h_{\mu\nu}=0
\,.
\label{fluct}
\ee
We decompose the general fluctuation into modes with definite mass
(from the $(D-1)$--dimensional point of view)
\be
h_{\mu\nu}(x^\si,y)
=
\sum\!\!\!\!\!\!\!\int
e^{i k_\si x^\si}h_m{}_{\mu\nu}(y)
\ee
where $k$ is a $(D-1)$--dimensional momentum satisfying $k^2=-m^2$ and
the sum is for discrete modes while the integral is for continuous
modes (in the above formula and in the next one the factor
$\exp(i k_\si x^\si)$
represents the sum of independent {\it real} fluctuations with a given
mass). Let us now concentrate on one mode
$h_{m\mu\nu}(y)$. Calculating the curvature tensor for the metric
\bea
g_{\mu\nu}(x^\si,y)
&\!\!\!=\!\!\!&
g_{\mu\nu}^0(x^\si,y)
+\eps\,e^{i k_\si x^\si}h_m{}_{\mu\nu}(y)
\,,
\nn\\
g_{DD}(x^\si,y)
&\!\!\!=\!\!\!&
g_{DD}^0(x^\si,y)
\,,
\eea
the equation of motion (\ref{eqofm}) will be expanded to the first
order in $\eps$.
For the background metric (\ref{ds2}) the Weyl
tensor vanishes therefore $C^{ab}$ is already of order $\eps$:
\be
C^{ab}=\eps(C_1)^{ab}+{\cal O}(\eps^2)
\,.
\ee
Thus, only terms up to the first order in $(C_1)^{ab}$ should be kept
in (\ref{eqofm}).
There is a second contribution to the linearized equations of
motion for $h_{\mu\nu}$ coming from the correction to $K^a$,
\be
K^a = (K_0)^a +\eps (K_1)^a +{\cal O}(\eps^2)
\,,
\ee
and in the product (\ref{eqofm}) also for this contribution
only the first power of $(K_1)^a$ should be kept.

Using the above $\eps$ expansion we can write the equation of motion
(\ref{eqofm}) up to terms linear in $\eps$ in the form:
\be
0=e\sum_n\ka_n
\frac{2^{n}(D-1-n)!}{(D-2)^n(D-1-2n)!}
\left((H_0^{(n)})_M^N +\eps (H_1^{(n)})_M^N
+\eps (G_1^{(n)})_M^N\right)
\label{EoM}
\ee
where we introduced the following tensors
(note that $H_0^{(n)}$ have different normalization than the
corresponding tensors $H^{(n)}$ of ref.\ \cite{KMO})
\bea
\left(H_0^{(n)}\right)_M^N
&\!\!\!=\!\!\!&
\left(\de_M^N\de_{M_1}^{N_1}\de_{M_2}^{N_2}
\cdots\de_{M_{n}}^{N_{n}}
\pm{\rm perm.}\right)
(K_0)^{M_1}_{N_1}(K_0)^{M_2}_{N_2}\cdots (K_0)^{M_{n}}_{N_{n}}
\,,
\label{hdef}\\
\left(H_1^{(n)}\right)_M^N
&\!\!\!=\!\!\!&
n\left(\de_M^N\de_{M_1}^{N_1}\de_{M_2}^{N_2}
\cdots\de_{M_{n}}^{N_{n}}
\pm{\rm perm.}\right)
(K_1)^{M_1}_{N_1}(K_0)^{M_2}_{N_2}\cdots (K_0)^{M_{n}}_{N_{n}}
\,,
\label{fdef}\\
\left(G_1^{(n)}\right)_M^N
&\!\!\!=\!\!\!&
\frac{n(D-2)}{4(D-1-n)}\cdot
\nn\\
&&
\cdot\left(\de_M^N\de_{M_0}^{N_0}\de_{M_1}^{N_1}
\cdots\de_{M_{n}}^{N_{n}}
\pm{\rm perm.}\right)
(C_1)^{M_0 M_1}_{N_0 N_1}(K_0)^{M_2}_{N_2}\cdots (K_0)^{M_{n}}_{N_{n}}
\,.\ \ \ \ \ \
\label{gdef}
\eea
In \cite{KMO} it was shown that the lowest order ($\eps^0$)
term vanishes for the domain wall metric (\ref{ds2})
with the warp factor $\si$ determined in terms of the
coupling constants $\ka_n$ (this domain wall is flat when some
relation among $\ka_n$ is satisfied).

Let us start the discussion of corrections of order
$\eps$ from $H_1^{(n)}$. From (\ref{fdef}) one can get the recurrence
relations for $H_1^{(n)}$
(valid not only for the domain wall but for all background metrics):
\be
(H_1^{(n)})_M^N
=
n\left(\de_M^N (K_1)_P^Q (H_0^{(n-1)})_Q^P
- (K_1)_M^P (H_0^{(n-1)})_P^N
- (K_0)_M^P (H_1^{(n-1)})_P^N\right).
\label{frec}
\ee
To proceed further let us specialize to the transverse and traceless
fluctuations of the metric (\ref{gfluct}) in the domain wall background
(\ref{ds2}). In such a case the tensor $K_1$ satisfies the following
conditions
\be
(K_1)^\mu_\mu=0,
\ \ \ \ \
(K_1)^D_\mu=(K_1)_D^\mu=(K_1)^D_D=0\,.
\label{K1}
\ee
Tensor $K_0$ has been calculated in \cite{KMO}:
\bea
(K_0^{(n)})_\mu^\nu
&\!\!\!=\!\!\!&
-\de_\mu^\nu \frac{D-2}{2}
\left(\frac{\pa f}{\pa y}\right)^2
\!,
\nn\\
(K_0^{(n)})^D_D
&\!\!\!=\!\!\!&
-\frac{D-2}{2}\left(\frac{\pa f}{\pa y}\right)^2
+\frac{\pa^2 f}{\pa y^2}
\,,
\label{K0}
\eea
while $H_0^{(n)}$ differ only by normalization from $H^{(n)}$
defined in \cite{KMO}:
\bea
(H_0^{(n)})_\mu^\nu
&\!\!\!=\!\!\!&
\de_\mu^\nu (-1)^n \left(\frac{D-2}{2}\right)^n
\frac{(D-1)!}{(D-1-n)!}
\cdot
\nn\\ \ \ \ \ \ \ \ \ \ \
&&\
\cdot
\left(\frac{\pa f}{\pa y}\right)^{2n-2}
\left[
\left(\frac{\pa f}{\pa y}\right)^2
- \frac{2n}{D-1}
\left(\frac{\pa^2 f}{\pa y^2}\right)
\right]
\,,\ \ \ \ \nn\\
(H_0^{(n)})^D_D
&\!\!\!=\!\!\!&
(-1)^n \left(\frac{D-2}{2}\right)^n
\frac{(D-1)!}{(D-1-n)!}
\left(\frac{\pa f}{\pa y}\right)^{2n}
\,.
\label{H0}
\eea
Substituting eqs.\ (\ref{K0}) and (\ref{H0}) into formula (\ref{frec})
and using the conditions (\ref{K1}) we get the following expression
for $H^{(n)}_1$
\be
(H_1^{(n)})_\mu^\nu
=
(K_1)_\mu^\nu\sum^n_{k=1}(-1)^k\frac{n!}{(n-k)!(D-1)^k}
\left((K_0)^\rho_\rho\right)^{k-1}
(H_0^{(n-k)})^\si_\si
\ee
with all other components vanishing. It explicitly reads
\bea
(H_1^{(n)})_\mu^\nu
=
(K_1)_\mu^\nu
&{}&\!\!\!\!\!\!\!\!
(-1)^n
\left(
\frac{D-2}{2}\right)^{n-1}
\frac{n(D-2)!}{(D-1-n)!}
\cdot
\nn\\
&&
\ \ \ \ \ \cdot
\left(
\frac{\pa f}{\pa y}\right)^{2n-4}
\left[
\left(\frac{\pa f}{\pa y}\right)^2
-
\frac{2(n-1)}{D-2}
\left(\frac{\pa^2 f}{\pa y^2}\right)
\right]
.\ \ \ \ \
\label{H1expl}
\eea

Let us now turn to $G_1^{(n)}$. For a general background metric
$g^0_{MN}$ the expression obtained by taking into account all
permutations in eq.\ (\ref{gdef}) is very complicated. The result
contains different combinations of all $H_0^{(k)}$ with $k<n$ and
it is not possible to write it as a simple recurrence analogous to
eq.\ (\ref{frec}) valid for $H_1^{(n)}$. Therefore, we will present
explicit formulae for $G_1^{(n)}$ only for the domain wall background
(\ref{ds2}). In such a case, using the symmetry properties of the Weyl
tensor $C$ and the fact that $(K_0)_\mu^\nu$ and $(H_0^{(k)})_\mu^\nu$
are proportional to $\de_\mu^\nu$, we get
\bea
(G_1^{(n)})_\mu^\nu
&\!\!\!=\!\!\!&
(C_1)_{\mu D}^{\nu D}
\,
\frac{n! (D-2)}{D-1-n}
\cdot
\nn\\
&&
\!\!\!\!\cdot
\sum_{k=1}^{n}
\frac{k(-1)^k}{(n-k)!(D-1)^k}
\left((K_0)^\rho_\rho\right)^{k-1}
\left((D-1)(H_0^{(n-k)})^D_D-(H_0^{(n-k)})^\rho_\rho\right)
\!.\nn\\
\eea
Substituting the explicit formulae for $K_0$ and $H_0^{(k)}$
(\ref{K0},\ref{H0}) it can be rewritten in the following form
\be
(G_1^{(n)})_\mu^\nu
=
(C_1)_{\mu D}^{\nu D}
(-1)^n \left(\frac{D-2}{2}\right)^{n-1}\!\!
\frac{2n(D-2)!}{(D-1-n)!}
\frac{(n-1)}{D-3}
\left(\frac{\pa^2f}{\pa y^2}\right)
\left(\frac{\pa f}{\pa y}\right)^{2n-4}
\!.
\label{G1expl}
\ee

We see that the tensors $H_1^{(n)}$ and $G_1^{(n)}$
are proportional to $K_1$ and $C_1$, respectively, wich can be
found using the decomposition of the curvature tensor
(given by eq.\ (\ref{eq:R})):
\bea
(K_1)_\mu^\nu
&\!\!\!=\!\!\!&
\left[
-\frac12\frac{\pa^2}{\pa y^2}
+\frac{D-5}{2}\left(\frac{\pa f}{\pa y}\right)\frac{\pa}{\pa y}
\right.
\nn\\
&&\hspace{16mm}
\left.
+(D-3)\left(\frac{\pa f}{\pa y}\right)^2
-\left(\frac{\pa^2 f}{\pa y^2}\right)
-\frac12m^2e^{2f}
\right]h_m{}_\mu^\nu(y)
\,,
\label{K1expl}\\
(C_1)_{\mu D}^{\nu D}
&\!\!\!=\!\!\!&
\frac{D-3}{D-2}
\left[
-\frac{1}{2}\frac{\pa^2}{\pa y^2}
-\frac{3}{2}\left(\frac{\pa f}{\pa y}\right)\frac{\pa}{\pa y}
\right.
\nn\\
&&\hspace{16mm}
\left.
-\left(\frac{\pa f}{\pa y}\right)^2
-\left(\frac{\pa^2 f}{\pa y^2}\right)
-\frac{1}{2(D-3)}m^2e^{2f}
\right]h_m{}_\mu^\nu(y)
\,.\ \ \ \
\label{C1expl}
\eea
In both these tensors the operators acting on $h_m{}_\mu^\nu(y)$ do
not depend on the Lorentz indices $\mu$, $\nu$ so we will drop these
indeces from now on.

Now we are ready to get the equation of motion for the fluctuation
$h_m(y)$. We rewrite equation (\ref{EoM}) using the explicite form of
$H_1^{(n)}$, $G_1^{(n)}$, $K_1$ and $C_1$
eqs.\ (\ref{H1expl},\ref{G1expl}--\ref{C1expl}).
The part of this equation linear in $\eps$ gives:
\bea
0
&\!\!\!=\!\!\!&
\sum_n^{n_{max}}
\ka_n(-1)^n\frac{2n(D-3)!}{(D-1-2n)!}\cdot
\nn\\
&&
\cdot\left[
-\frac{1}{2}\left(f'\right)^{2n-2}
\frac{\pa^2}{\pa y^2}
+\frac{D-5}{2}\left(f'\right)^{2n-1}
\frac{\pa}{\pa y}
-(n-1)\left(f'\right)^{2n-3}
\left(f''\right)
\frac{\pa}{\pa y}
\right.\ \ \
\nn\\
&&
\left.\,\,\,\,\,
+(D-3)\left(f'\right)^{2n}
-(2n-1)\left(f'\right)^{2n-2}
\left(f''\right)
-\frac{1}{2}m^2e^{2f}\left(f'\right)^{2n-2}
\right.
\nn\\
&&
\left.\,\,\,\,\,
+\frac{n-1}{D-3}m^2e^{2f}\left(f'\right)^{2n-4}\left(f''\right)
\right]h_m(y)
\,.
\label{EoMexpl}
\eea
At this point we can use the explicit form of the warp factor
$f(y)=\si|y|$. One should be careful when performing this substitution
because the first derivative of $f(y)$ is not continuous at $y=0$ and
the second derivative of $f(y)$ is proportional to the Dirac delta at
$y=0$. Similar care is needed when calculating terms containing
derivatives of the fluctuation $h_m$ because the solutions to the
above equation are also functions of $|y|$. Regularizing the Dirac
delta function one can find the following equalities\footnote{
The necessary denominators of $2k+1$ have not been taken into
account for example by the authors of ref.\ \cite{KKL} (who discuss the
case $D=5$) leading to wrong equations of motion.}:
\bea
(f')^{2k}(f'')
&\!\!\!=\!\!\!&
\frac{2}{2k+1}\si^{2k+1}\de(y)
\,,\\
(f')^{2k}\frac{\pa^2}{\pa y^2}h_m(y)
&\!\!\!=\!\!\!&
\si^{2k}h_m''(|y|)+\frac{2}{2k+1}\si^{2k}\de(y)h_m'(0^+)
\,,\\
(f')^{2k-1}(f'')\frac{\pa}{\pa y}h_m(y)
&\!\!\!=\!\!\!&
\frac{2}{2k+1}\si^{2k}\de(y)h_m'(0^+)
\eea
where
\be
h_m'(0^+)=\lim_{y\to0^+}\frac{\pa}{\pa y}h_m(y)
\,.
\ee
Making these substitutions in eq.\ (\ref{EoMexpl}) we find the
following bulk ($y\ne 0$) equation of motion for $h_m$ :
\bea
0
&\!\!\!=\!\!\!&
\sum_n^{m_{max}}
\ka_n(-1)^n\frac{n(D-3)!}{(D-1-2n)!}\si^{2n-2}\cdot
\nn\\
&&\ \ \ \ \ \ \cdot
\left[
-h_m'' +(D-5)\si h_m' +2(D-3)\si^2 h_m-m^2e^{2\si|y|}h_m
\right]
.\ \ \
\label{EoMbulk}
\eea
It is rather amazing that the expression in the square parenthesis does
not depend on $n$. Therefore for arbitrary $\ka_n$ and $n_{max}$ the bulk
equation of motion for $h_m$ reduces to
\be
0=-h_m'' +(D-5)\si h_m' +2(D-3)\si^2 h_m-m^2e^{2\si|y|}h_m
\,.
\ee
Its solution for $m=0$ is equal to
\be
h_0(y)=A_0e^{-2\si|y|}+B_0e^{(D-3)\si|y|}
v\label{sol0}
\ee
while for $m\ne0$ it can be written using the Bessel functions of
order $\frac{D-1}{2}$:
\be
h_m(y)
=
e^{\left(\frac{D-5}{2}\si|y|\right)}
\left[
A_m J_\frac{D-1}{2}\left(\frac{m}{\si}e^{\si|y|}\right)
+B_m Y_\frac{D-1}{2}\left(\frac{m}{\si}e^{\si|y|}\right)
\right].
\label{solm}
\ee
For each $m$ only one of the above combinations of the solutions of
the bulk equation (\ref{EoMbulk}) is the solution of the full equation
(\ref{EoMexpl}). In order to identify this combination we have to take
into account the part of the equation of motion (\ref{EoMexpl})
proportional to $\de(y)$. It reads:
\bea
0
&\!\!\!=\!\!\!&
\sum_n^{n_{max}}
\ka_n(-1)^n\frac{n(D-3)!}{(D-1-2n)!}\si^{2n-1}\cdot
\nn\\
&&\ \ \ \ \ \cdot
\left[
-\frac{h_m'(0^+)}{\si} - 2h_m(0)
+\frac{2(n-1)}{(D-3)(2n-3)}\frac{m^2}{\si^2}h_m(0)
\right]
.\ \ \
\label{dely}
\eea
This equation is equivalent to the following boundary condition at
$y=0$:
\be
\frac{h_m'(0^+)}{h_m(0)}=-2\si
\left(
1-\frac{m^2}{\si^2}\frac{1}{D-3}
\frac
{\sum_n\ka_n(-1)^n\frac{n}{(D-1-2n)!}\frac{(n-1)}{(2n-3)}\si^{2n}}
{\sum_n\ka_n(-1)^n\frac{n}{(D-1-2n)!}\si^{2n}}
\right).
\ee
For $m\ne0$ it is quite complicated and depends on $\ka_n$ but in
principle can be used to fix the ratio of the coefficients $A_m/B_m$
for the massive modes.

The situation is much simpler for $m=0$ because in this case the
above boundary condition does not depend on $\ka_n$ and simplifies to
\be
\frac{h_0'(0^+)}{h_0(0)}=-2\si
\,.
\ee
Applying this condition to (\ref{sol0}) we find that
for arbitrary space--time dimension, $D$, and for arbitrary strength
of the higher order interactions (given by the coefficients
$\ka_n$) the massless solution of the equations of motion always exists
and is given by
\be
h_0(y)=\exp\left(-2\si|y|\right)
.
\ee

We would like to interpret these solutions as massless, normalizable
4--dimensional gravitons. It turns out to be possible, but  some
care is needed. Let us
start with the normalizability issue. To check whether the above solutions
(massless and massive) are normalizable we have to choose an
appropriate integration measure. Such measure can be determined by the
requirement that the kinetic operator for the gravitons should be
self--adjoint (which is not fulfilled in the case of eq.\
(\ref{EoMbulk})). To find the proper operator it is helpful to change the
variable $y$ and to rescale fluctuation $h_m$:
\be
h_m(y)=\left(1+\si|z|\right)^{\frac{D-6}{2}} \hat h_m(z)
\ee
where
\be
\si z = {\rm sgn}(y)\left(e^{\si|y|}-1\right)
.
\ee

The equation of motion reads then
\be
-\frac{d^2}{dz^2}h_m(z)+\frac{D(D-2)\si^2}{4(1+\si |z|)^2}h_m(z)=m^2 h_m(z)
\ee
and is explicitly self--adjoint with a flat measure (which we can take equal
to 1).

The properly normalized  massless solution is given by
\be
\hat h_0(z)=\sqrt{\frac{(D-3)\si}{2}}\left(1+\si|z|\right)^{-\frac{D-2}{2}}
\ee
and is normalizable (for $D>3$) while the massive modes are given by
the formula
\bea
\hat h_m(z)
&=&\sqrt{\frac{m}{\si}+m|z|}\cdot\\
&&\ \ \ \cdot\left[
A_m J_{\frac{D-1}{2}}\left(\frac{m}{\si}+m|z|\right)
+B_m \left(\frac{m}{\si}\right)^{D-3}
Y_{\frac{D-1}{2}}\left(\frac{m}{\si}+m|z|\right)
\right]\nn
\eea
Massive modes are asymptotically (for large $|z|$) plane waves and
therefore for infinite range of $z$ not normalizable.

We showed that the massless mode is always normalizable but this is
not enough to interpret it as a graviton. The reason is the
following. The bulk equation of motion (\ref{EoMexpl}) contains an
overall factor depending on $\ka_n$ and on the warp factor $\si$:
\be
\sum_n^{n_{max}}
\ka_n(-1)^{(n-1)}\frac{n(D-3)!}{(D-1-2n)!}\si^{2n-2}
\,.
\label{coeff}
\ee
Although the value of this factor is not important for the solution,
we have to remember that
the sign of this factor is related to the sign
of the kinetic energy term for the fluctuations in the effective
lagrangian. The wrong sign of the kinetic energy indicates instability
of the assumed background. Thus, our domain wall solution can be
stable only if the parameters $\ka_n$ are such that the sum in eq.\
(\ref{coeff}) is positive. Using rescaled couplings (introduced in
\cite{KMO}):
\be
p_n=(-1)^{n-1}\ka_n\frac{(D-1)!}{(D-1-2n)!}
\ee
the necessary condition for the stability of the domain wall solutions
can be therefore written in the form
\be
\sum_{n=1}^{n_{max}} np_n\si^{2n-2}>0
\,.
\label{sign}
\ee

It is interesting to compare the above condition with the bulk and
boundary equations \cite{KMO} which must be satisfied by the warp
factor, $\si$, of the background domain wall metric (\ref{ds2}):
\bea
\sum_{n=1}^{n_{max}} p_n \si^{2n}
&\!\!\!=\!\!\!&
-\La
\,,
\label{blcond}
\\
\sum_{n=1}^{n_{max}} \frac{n}{2n-1} p_n \si^{2n-1}
&\!\!\!=\!\!\!&
\frac{D-1}{4}\la
\,.
\label{llcond}
\eea
In our previous paper \cite{KMO} we have considered possibility of
the domain wall solutions without the bulk or/and the brane
cosmological constants. Such solutions are acceptable only if the new
condition (\ref{sign}) is satisfied and this must be checked for any
specific model described by a set of the coefficients $\ka_n$.
It is not easy to  discuss the consequences of
(\ref{sign}) in general but one can make the following observation.

Let us assume that we insist on $\la=0$ -- the ''self--supporting'' brane
solution without any matter on the brane (the bulk cosmological constant
may be different from 0). Let us define a polynomial
\be
P(x)=\sum_{n=1}^{n_{max}}
\frac{np_n}{2n-1} x^{2n-1}\,.
\label{pol}
\ee
The condition for vanishing $\la$ reads
$P(\si)=0$
while the condition for the correct sign of the kinetic terms is
$P'(\si)>0$. From the positivity of the gravitational constant
$\ka_1=(2\ka^2)^{-1}$ it follows that $P'(0)>0$. This means of course
that the trivial domain wall with $\si=0$ (i.e. the Minkowski space)
has the correct sign of the graviton kinetic term. This means also
that the first nontrivial domain wall (the one with the smallest
positive $\si$) with $\la=0$ is unstable. But not all solutions with
$\la=0$ must be unstable. The number of different solutions with
$\la=0$ and $\si \ne 0$ is equal to the number of positive zeros of
$P(x)$ which is at most $(n_{max}-1)$.
If $P(x)$ has only first order zeros than ``every second'' solution
with $\la=0$ can be stable because such function changes is derivative
when moving from one zero point to the next one.
Thus, the nontrivial domain wall with vanishing brane cosmological
constant is possible if $n_{max} \ge 3$ and this can be satisfied
for space--time dimension $D \ge 6$.

Similar analysis can be performed for domain walls with vanishing bulk
cosmological constant, $\La=0$ -- the only difference is to use the 
polynomial $\sum_{n=1}^{n_{max}}p_n x^{2n}$ instead of that defined in 
eq. (\ref{pol}).

In the case of both cosmological constants vanishing it is not difficult to 
see that for $n_{max}\ge 3$ there exists a range
of values of $p_1,\ldots,p_{n_{max}}$ for which all three conditions
(\ref{sign}--\ref{llcond}) can be simultaneously satisfied
with $\la=\La=0$ and $\si\ne 0$. Thus, stable solutions with
vanishing brane and/or bulk cosmological constant are possible if the
space--time is at least 6--dimensional.

In conclusion,
we considered a class of models with higher order gravity corrections
in the form of the Euler densities with arbitrary power $n$ of the
curvature tensor in arbitrary space--time dimension $D$.
The fluctuations around the domain wall type solutions (found in \cite{KMO})
were shown to have similar spectrum as in the lowest order case
($n=1$) -- the bulk equation of motion rather miraculously turned out to
depend only on $D$ and not on $n$. The boundary condition at the
wall for massive modes has some $n$--dependence.
There exists one normalizable massless mode and a continuum
of massive modes (without the energy gap).
The solutions for all $D$ are almost the same as in the original
Randall--Sundrum model with the Hilbert--Einstein
action \cite{RS2} (apart from
some numerical factors) and the discussion about the applicability of
the Newton's law and the effective number of dimensions can be carried
over from (\cite{RS2}) to the general case discussed in this paper virtually
unchanged.

\vspace{1cm}

This work was partially supported by the Polish KBN grants
5 P03B 150 20, 2 P03B 052 16
and by the European Commission RTN grant HPRN--CT--2000--00152.

\end{document}